\documentclass[twocolumn]{aastex61}

 \usepackage{longtable}
 \usepackage{rotating}
 \usepackage{multirow}

\begin{document}

\def\liu#1{{\bf #1}}

\shorttitle{On the Transition of the Galaxy Quenching Mode at $0.5<\lowercase{z}<1$}
\title{On the Transition of the Galaxy Quenching Mode at $0.5<\lowercase{z}<1$ in CANDELS
}
\correspondingauthor{F. S. Liu}
\email{Email: fengshan.liu@yahoo.com}

\author{F. S. Liu $^{\color{blue} \dagger}$}
\affil{College of Physical Science and Technology, Shenyang Normal University, Shenyang 110034, China}
\affiliation{University of California Observatories and the Department of Astronomy and Astrophysics,
University of California, Santa Cruz, CA 95064, USA}

\author{Meng Jia}
\affil{College of Physical Science and Technology, Shenyang Normal University, Shenyang 110034, China}

\author{Hassen M. Yesuf}
\affiliation{University of California Observatories and the Department of Astronomy and Astrophysics,
University of California, Santa Cruz, CA 95064, USA}

\author{S. M. Faber}
\affiliation{University of California Observatories and the Department of Astronomy and Astrophysics,
University of California, Santa Cruz, CA 95064, USA}

\author{David C. Koo}
\affiliation{University of California Observatories and the Department of Astronomy and Astrophysics,
University of California, Santa Cruz, CA 95064, USA}

\author{Yicheng Guo}
\affiliation{Department of Physics and Astronomy, University of Missouri, Columbia, MO 65211, USA}

\author{Eric F. Bell}
\affiliation{Department of Astronomy, University of Michigan, Ann Arbor, MI, USA}

\author{Dongfei Jiang}
\affil{College of Physical Science and Technology, Shenyang Normal University, Shenyang 110034, China}

\author{Weichen Wang}
\affiliation{Department of Physics \& Astronomy, Johns Hopkins University, 3400 N. Charles Street, Baltimore, MD 21218, USA}

\author{Anton M. Koekemoer}
\affil{Space Telescope Science Institute, 3700 San Martin Drive, Baltimore, MD 21218, USA}

\author{Xianzhong Zheng}
\affiliation{Purple Mountain Observatory, Chinese Academy of Sciences,
2 West-Beijing Road, Nanjing 210008, China}

\author{Jerome J. Fang}
\affiliation{Astronomy Department, Orange Coast College, Costa Mesa, CA 92626, USA}

\author{Guillermo Barro}
\affiliation{University of the Pacific, Department of Physics, 3601 Pacific Avenue Stockton, CA 95211, USA}

\author{Pablo G. P\'erez-Gonz\'alez}
\affiliation{Departamento de Astrof\'isica, Facultad de CC. F\'isicas, Universidad Complutense de Madrid, E-28040 Madrid, Spain}

\author{Avishai Dekel}
\affil{Center for Astrophysics and Planetary Science, Racah Institute of Physics, The Hebrew University, Jerusalem, Israel}

\author{Dale Kocevski}
\affil{Department of Physics and Astronomy, Colby College, Mayflower Hill Drive, Waterville, ME 0490, USA}

\author{Nimish P. Hathi}
\affil{Space Telescope Science Institute, 3700 San Martin Drive, Baltimore, MD 21218, USA}

\author{Darren Croton}
\affil{Centre for Astrophysics \& Supercomputing, Swinburne University of Technology, P.O. Box 218, Hawthorn, Victoria 3122, Australia}

\author{M. Huertas-Company}
\affil{Sorbonne Université, Observatoire de Paris, Université PSL, Université Paris Diderot, CNRS, LERMA, Paris, France} 

\author{Xianmin Meng}
\affil{National Astronomical Observatories, Chinese Academy of Sciences, A20 Datun Road, Beijing 100012, China}

\author{Wei Tong}
\affil{College of Physical Science and Technology, Shenyang Normal University, Shenyang 110034, China}

\author{Lu Liu}
\affil{College of Physical Science and Technology, Shenyang Normal University, Shenyang 110034, China}

%\author{et al.}

\begin{abstract}

We investigate the galaxy quenching process at intermediate redshift using a sample of
$\sim4400$ galaxies
with $M_{\ast} > 10^{9}M_{\odot}$ between redshift 0.5 and 1.0 in all five CANDELS fields. 
We divide this sample, using the integrated specific star formation rate (sSFR), 
into four sub-groups: star-forming galaxies (SFGs) above and below the ridge of 
the star-forming main sequence (SFMS), transition galaxies and quiescent galaxies. 
We study their $UVI$ ($U-V$ versus $V-I$) color gradients to infer their sSFR gradients 
out to twice effective radii. 
We show that on average both star-forming and transition galaxies at all masses 
are not fully quenched at any radii, whereas quiescent galaxies are fully quenched at all radii.  
We find that at low masses ($M_{\ast} = 10^{9}-10^{10}M_{\odot}$) SFGs both above and 
below the SFMS ridge generally have flat sSFR profiles, 
whereas the transition galaxies at the same masses generally have sSFRs that are more 
suppressed in their outskirts. In contrast, at high masses ($M_{\ast} > 10^{10.5}M_{\odot}$), 
SFGs above and below the SFMS ridge and transition galaxies generally 
have varying degrees of more centrally-suppressed sSFRs relative to their outskirts.
These findings indicate that at $z\sim~0.5-1.0$ the main galaxy quenching mode depends on 
its already formed stellar mass, exhibiting a transition from ``the outside-in'' 
at $M_{\ast} \leq 10^{10}M_{\odot}$ to ``the inside-out'' at $M_{\ast} > 10^{10.5}M_{\odot}$. 
In other words, our findings support that internal processes 
dominate the quenching of massive galaxies, whereas external processes dominate the quenching of low-mass galaxies.


\end{abstract}

\section{Introduction}

Studying the spatial distribution of specific star formation rate ($sSFR= SFR/M_{\ast}$) 
is helpful to understand how stellar mass ($M_{\ast}$) is built up in galaxies as they 
evolve along the star formation main sequence \citep[SFMS,][]{Brinchmann2014,Noeske2007,Elbaz2007,Whitaker14}, 
and how and where the star formation shuts down as galaxies move off the SFMS to become fully quiescent. 
Broadly speaking, there are two types of processes responsible for cessation of star formation in galaxies: 
the internal and the external processes. The internal processes 
quench star formation due to the intrinsic properties of galaxies, such
as central compaction \citep[][]{Fang2013,Liu16,Barro2017,Whitaker17},
AGN feedback \citep[][]{Croton2006} and supernova feedback \citep[][]{Geach2014}, which 
scale with the stellar mass of galaxies \citep[``mass-quenching''; e.g.,][]{P10}. 
The internal processes first deplete the gas in the centers of galaxies or blow it out of the centers, 
causing ``the inside-out'' quenching. 
The external processes, like the environmental effects \citep[e.g.,][]{P12,Geha2012,Guo2017}, 
strip the gas content of galaxies first from their outskirts, 
causing ``the outside-in'' quenching. The two types of processes 
are expected to change the radial sSFR profile of a galaxy in a different way 
during its quenching process.

Encouraging progress has recently been made in understanding the radial gradients of 
sSFR, traced by $EW({\rm H{\alpha}})$ and rest-frame UV-optical colors, 
in distant star-forming galaxies (SFGs) \citep[e.g.,][]{Wuyts12,Nelson12,Nelson16b,Nelson16a, 
Liu16,Liu17,Wang2017,Tacchella17}). These studies typically find either flat sSFR gradients 
(in galaxies with $M_{\ast} \lesssim 10^{10}M_{\odot}$ at $z\sim1$ and galaxies 
with $M_{\ast} < 10^{11}M_{\odot}$ at $z\sim2$) 
or somewhat centrally-suppressed sSFRs (in galaxies 
with $M_{\ast} \gtrsim 10^{10.5}M_{\odot}$ at $z\sim1$ and galaxies 
with $M_{\ast} > 10^{11}M_{\odot}$ at $z\sim2$). 
A correction for dust gradient was shown to be one of the main sources
of uncertainty that makes studying the sSFR gradients in distant SFGs 
challenging \citep[][]{Wuyts2013,Liu16,Liu17,Wang2017,Tacchella17,Nelson2018}.

Rest-frame $UVJ$ ($U-V$ versus $V-J$) diagram has been widely used 
to separate quenched from dusty/star-forming galaxies 
\citep[e.g.,][]{Wuyts2007,Williams09,Patel11,Brammer2011}. 
More recently, it has also been successfully utilized to determine sSFR and $A_V$ values, 
which are broadly consistent with 
the values derived from fitting reddened stellar population models to 
broad-band spectral energy distributions (SEDs) of galaxies covering 
UV to mid-infrared \citep[][]{Fang2017}. 
Furthermore, \citet[][]{Wang2017} demonstrated that rest-frame $UVI$ ($U-V$ versus $V-I$) diagram is 
as useful as the $UVJ$ diagram for distinguishing sSFR from dust extinction.

In cosmological simulations, \citet[][]{Tacchella06a,Tacchella06b} predict that, at high redshifts ($z=7-1$), 
the galaxy evolution across the SFMS is associated with events of wet compaction into compact 
star-forming systems, which trigger central gas depletion and the formation of an extended 
gas ring around it. The Tacchella et al. simulations reveal that the high-sSFR galaxies at the upper envelope 
of the SFMS and the lower-sSFR galaxies at the lower envelope of the SFMS 
have different properties, which is closely related to the quenching process of galaxies. 

With the aim to understand the star formation quenching process in distant galaxies,
in this work we select a sample of 4377 galaxies
with $M_{\ast} > 10^{9}M_{\odot}$ between redshift 0.5 and 1.0 from
all five CANDELS fields. 
With this sample, we investigate the $UVI$ color gradients and inferred
sSFR gradients in various galaxy populations (i.e., star-forming galaxies, transition galaxies and 
quiescent galaxies). We specifically follow \citet[][]{Tacchella06a,Tacchella06b} to divide our 
SFGs at intermediate redshifts into above and below the ridge of the SFMS, since this further classification 
is likely helpful to shed light on whether the shape of sSFR profiles start to vary during the evolution 
across the SFMS. We show that these different populations of galaxies have varying degrees of color and sSFR gradients.
We find that the main quenching mode of a galaxy at $z \sim 0.5-1$
depends on its already formed stellar mass, and it is outside-in for galaxies
with $M_{\ast} \leq 10^{10}M_{\odot}$, and inside-out for galaxies with $M_{\ast} > 10^{10.5}M_{\odot}$.
Throughout the paper, we adopt a cosmology with a matter density 
parameter $\Omega_{\rm m}=0.3$, a cosmological constant 
$\Omega_{\rm \Lambda}=0.7$ and a Hubble constant of ${\rm H}_{\rm 0}=70\,{\rm km \, s^{-1} Mpc^{-1}}$. 
All magnitudes are in the AB system.

\section{DATA}
CANDELS \citep[][]{Grogin11,Koekemoer11} is an
{\it HST} Multi-Cycle Treasury Program to image portions of five commonly studied
legacy fields (COSMOS, EGS, GOODS-N, GOODS-S and UDS).
The CANDELS group has made a multi-wavelength photometry catalog
for each field. Photometry in {\it HST}/WFC3 and ACS was measured
by running {\tt SExtractor} in dual model on the point spread function (PSF)-matched
images, with the F160W image as the detection image.
Photometry in the lower-resolution images (e.g., ground-based and IRAC)
was measured using {\tt TFIT} \citep[][]{Laidler07}. We refer readers to
\citet[][]{Guo+13}, \citet[][]{Galametz13}, \citet[][]{Nayyeri2017}, \citet[][]{Stefanon2017},
and Barro et al. (in preparation) for details.

Redshifts used in this work are in the priority order of secure spectroscopic (flagged as ``very secure'' or ``reliable''), 
good grism (at least two users agree that it is good) and photometric redshifts if available. 
Spectroscopic redshifts were recently re-compiled by N. P. Hathi (private communication) for all five CANDELS fields, 
which include publicly available \citep[e.g.,][and reference therein]{Santini+15} and 
unpublic (e.g., UCR DEIMOS Survey) redshifts. 
Grism redshifts came from the 3D-HST/CANDELS Survey \citep[e.g.,][]{Morris15,Momcheva16}. 
Photometric redshifts were estimated using the multi-wavelength
photometry catalogs and adopting a hierarchical Bayesian approach \citep[][]{Dahlen13}. 
The typical scatter of photometric redshifts spans from 0.03 to 0.06 in $\Delta(z)/(1+z)$.
Rest-frame integrated magnitudes from $FUV$ to $K$ 
were computed using {\tt EAZY} \citep[][]{Brammer+08}, which 
fits a set of galaxy SED templates to the multi-wavelength photometry, with the redshifts as inputs.

Stellar masses were computed using {\tt FAST} \citep[][]{Kriek09} and 
based on a grid of \citet[][]{BC03} models
that assume a \citet[][]{Chabrier03} IMF, declining $\tau$-models, 
solar metallicity and a \citet[][]{Calzetti00} dust law. 
The typical formal uncertainty in stellar mass is $\sim 0.1$ dex. 
SFRs were computed from rest-frame $UV$ luminosities at $\lambda\approx2800\AA$ 
that are corrected for extinction by applying a foreground-screen Calzetti reddening 
law ($\rm A_{2800}\approx1.79A_V$):$ SFR_{\rm UV,cor}[M_{\odot}{yr}^{-1}]=2.59\times10^{-10}L_{\rm UV,cor}[L_{\odot}]$ 
\citep[][]{Kennicutt12}. 
\citet[][]{Fang2017} showed that the sSFRs by this method 
are consistent with those derived from UV and far-IR luminosities in 
a broad range, with typical scatter of $\sim0.2$ dex.   
We adopted the median $A_{\rm V}$ that was calculated by combining results from four methods 
\citep[see labeled $2a_{\tau}$, $12a$, $13a_{\tau}$ and $14a$ in ][]{Santini+15} 
if available. These methods were chosen based on the same assumptions 
(Chabrier IMF and the Calzetti dust law). 
The typical formal uncertainty in $A_{\rm V}$ is $\sim0.1$ mag.
Effective radius along the semi-major axis ($R_{\rm SMA}$) and minor-to-major axis ratio ($q$) were measured
from the F125W images using {\tt GALFIT} \citep[][]{Peng+02}
by \citet[][]{vdWel+12}.

Spatially-resolved data is taken from the {\it HST}-based
multi-band and multi-aperture photometry catalogs of CANDELS still
under construction by Liu et al. (in preparation). 
These datasets include the radial profiles of isophotal ellipticity ($\varepsilon$) and
disky/boxy parameter $\rm A_4$ in both F125W and F160W, and
the observed surface brightness profiles in all {\it HST}/ACS (F435W, F606W, F775W, F814W and F850LP) 
bands and WFC3 (F105W, F125W, F140W and F160W) bands if available. 
Preliminary imaging reduction prior to multi-aperture photometry can be found in \cite{Jiang2018}. 
The photometry was done by using the {\tt IRAF} routine {\tt ellipse} 
within STSDAS, which is based on a technique described by \cite{Jedrzejewski1987}. 
For galaxies used in this work, we fixed the galaxy geometric centers, ellipticities and position angles                
obtained from the {\tt GALFIT} measurements along the semi-major axes for all available bands. 
Rest-frame $U$, $V$ and $I$ band surface brightness profiles 
were then computed using {\tt EAZY} \citep[][]{Brammer+08} by fitting the best-fit SEDs 
in each photometry annulus \citep[refer to Figure 2 in][]{Liu16}.

\section{Sample Selection}

%%-------------
\begin{table*}[ht] 
\footnotesize
\begin{center}
\caption{Sample selection criteria and the resulting sample sizes for each field.}
\begin{tabular}{lcccccc}
\hline\hline
{\footnotesize Criterion} & {\footnotesize GOODS-S} &
{\footnotesize UDS} & {\footnotesize GOODS-N} & {\footnotesize EGS} & {\footnotesize COSMOS}
& {\footnotesize Combined}\\
\hline
Full catalog                  &  34930(100\%)   & 35932(100\%)    & ~35445 (100\%)   &  ~41457(100\%)   & ~38671(100\%)    & 186435(100\%) \\
Hmag $<$ 24.5                 &  ~8293(23.74\%)  & ~9671(26.91\%)   & ~9460(26.69\%)   &  ~~~11292(27.24\%) & ~~~11811(30.54\%)  &  ~~50527(27.10\%) \\
PhotFlag=0                    &  ~8104(23.20\%)  & ~9151(25.47\%)   & ~9011(25.42\%)   &  ~7521(18.14\%)  &  ~7603(19.66\%)  &  ~~41084(22.04\%) \\
CLASS$\_$STAR$<$0.9           &  ~7901(22.62\%)  & ~8952(24.91\%)   & ~8815(24.87\%)   &  ~7252(17.49\%)  &  ~7297(18.86\%)  &  ~~40217(21.57\%) \\
$0.5 < z < 1.0$               &  2460(7.04\%)   & 2331(6.49\%)    & 2746(7.75\%)    &  1950(4.70\%)   &  2457(6.35\%)   &  11944(6.41\%) \\
$log M/M_{\odot} > 9.0$       &  1291(3.70\%)   & 1293(3.60\%)    & 1632(4.60\%)    &   952(2.30\%)~   &  1428(3.69\%)   &   ~6596(3.54\%) \\
{GALFIT} flag(J) = 0 or 1     &  1232(3.53\%)   & 1246(3.47\%)    & 1572(4.44\%)    &   933(2.25\%)~   &  1373(3.55\%)   &   ~6356(3.41\%) \\
$\rm R_{SMA}>0.18{\arcsec}$   &  1092(3.13\%)   & 1089(3.03\%)    & 1369(3.86\%)    &   814(1.96\%)~   &  1181(3.05\%)   &   ~5545(2.97\%) \\
Accurate multi-band SBPs      &   963(2.76\%)~   &  828(2.30\%)~    & 1210(3.41\%)    &   681(1.64\%)~   &   695(1.80\%)   &   ~4377(2.38\%) \\
\hline
\end{tabular}
\end{center}
\end{table*}

In order to maximize the sample size, we select galaxies from all five CANDELS fields 
by applying the following criteria to the above catalogs: \\

\begin{enumerate}

\item  Observed F160W($H$) magnitude brighter than 24.5 to ensure 
high signal-to-noise ratios (S/Ns). \\ 

\item  {\tt SExtractor} $\rm PhotFlag = 0$ and $\rm CLASS\_STAR < 0.9$
to exclude spurious sources and stars. \\

\item  Redshifts within $0.5<z<1$ and stellar masses $\rm M_{\ast} > 10^{9}M_{\odot}$
to maintain high mass completeness and to guarantee the accuracy of rest-frame $U$, $V$ and
$I$ band spatially-revolved data. Note that the {\it HST}
imaging in CANDELS ends at observed $H$ band, which roughly
corresponds to the rest-frame $I$ band for galaxies at $z=1$. \\

\item {\tt GALFIT} quality $\rm flag = 0$ (good fit) or $\rm flag = 1$ (suspicious fit) 
in F125W \citep[][]{vdWel+12} to ensure well-constrained measurements 
of structural parameters (i.e., effective radius 
and minor-to-major axis ratio) and eliminate mergers and disturbed objects. \\

\item  $R_{\rm SMA}>0.18^{\prime\prime}~(\rm 3~drizzled~pixels)$ to minimize the PSF effects 
on color gradient measurement. This lack of sample completeness cannot be avoided for this study 
given the limited resolution of {\it HST} imaging (see Appendix). \\

\item  Accurate measurements of the surface brightness profiles (SBPs) 
from center to $2R_{\rm SMA}$ in at least two ACS bands 
and two WFC3 bands simultaneously to guarantee the accuracy of SED modelling 
in each photometry annulus. 
Almost all ($\sim98\%$) of galaxies after this cut 
have accurate multi-aperture photometry in $F606W$, $F814W$, $F125W$ and $F160W$, 
which cover all three rest-frame U, V and I bands at $0.5<z<1$. \\ 

\end{enumerate}

\begin{figure*}
\centering
\includegraphics[angle=0,width=0.85\textwidth]{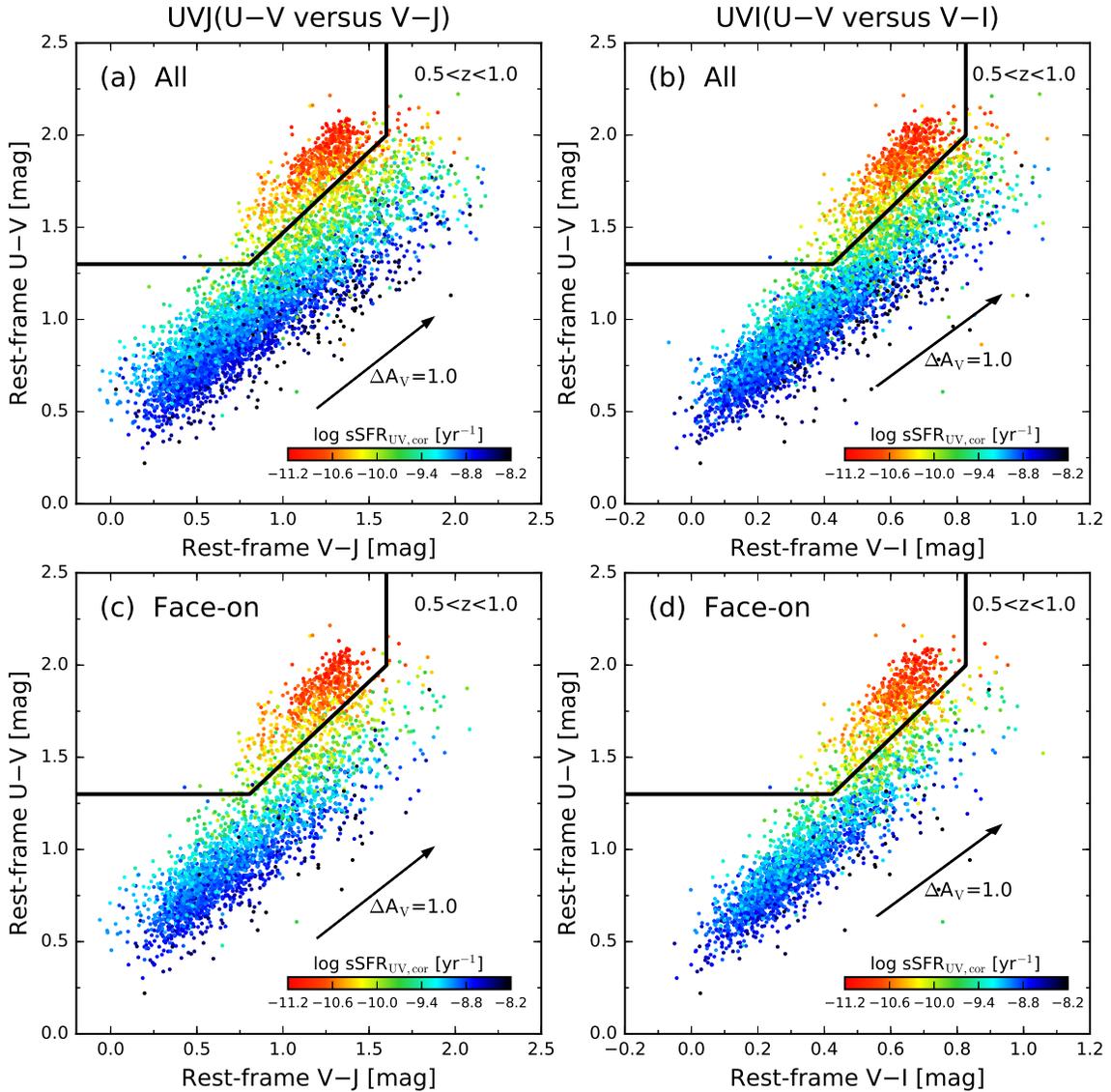}
\caption{
Rest-frame global $UVJ$ diagrams (left) and $UVI$ diagrams (right) for
our total sample (top) and the face-on ($q>0.5$) subsample
(bottom) after applying the selection criteria 1-5, respectively.
Data points are color-coded by $\rm log~sSFR_{UV,cor}$.
Solid lines in $UVJ$ diagrams indicate the boundary of \citet[][]{Williams09}
to separate quiescent from star-forming galaxies.
For $UVI$, the boundary lines are determined by visually confirming
that objects on the $UVJ$ boundaries also lie on the boundary lines in $UVI$.
The arrows indicate the Calzetti reddening vector.
\label{calibration}}
\end{figure*}

\begin{figure*}
\centering
\includegraphics[angle=0,width=0.8\textwidth]{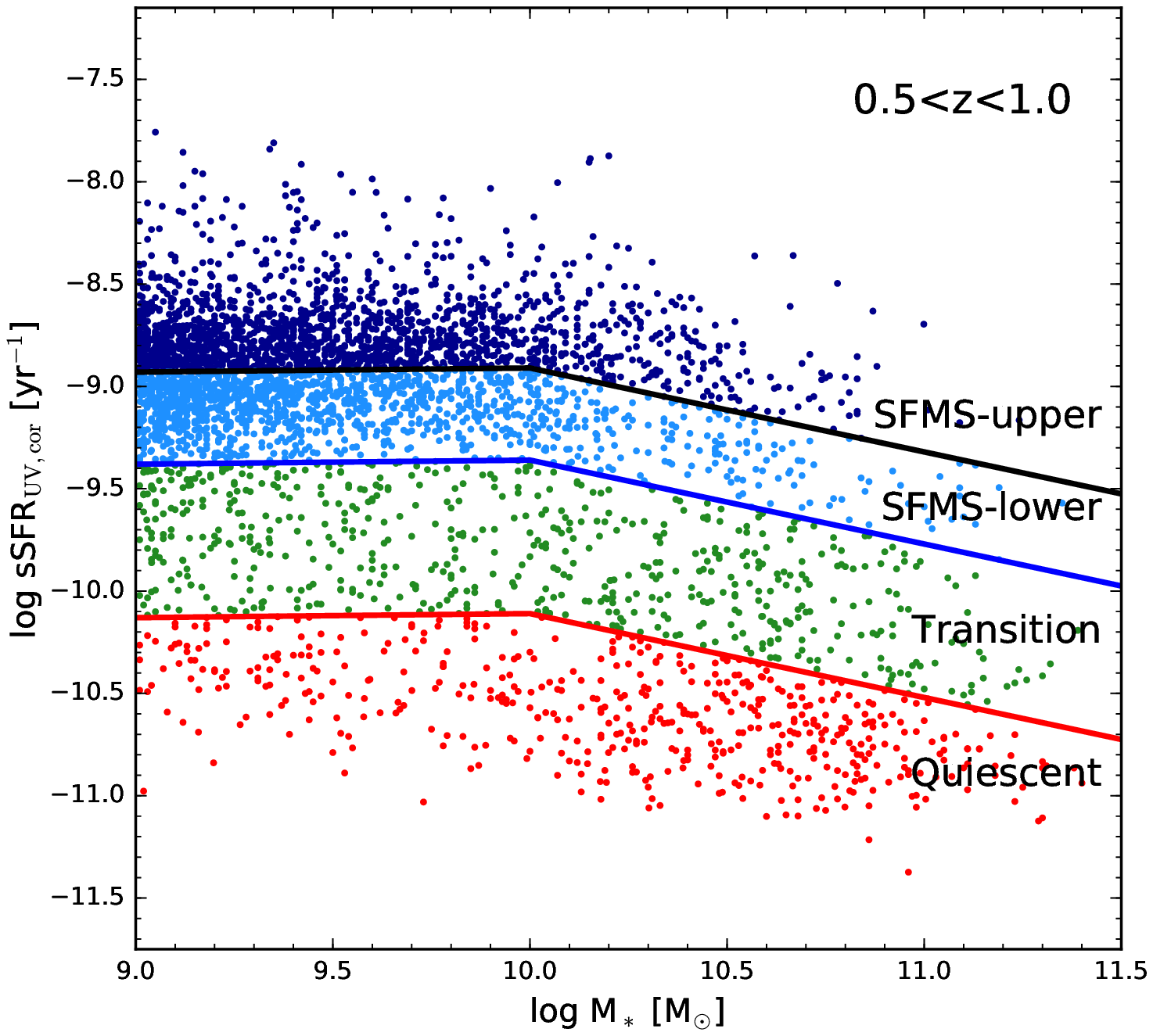}
\caption{
The sSFR-mass relation for our final sample of galaxies. 
Black lines are the best broken power-law fits to the SFMS (namely $\rm {\Delta}~log~sSFR_{\rm UV,cor}=0$). 
Blue lines indicate $\rm {\Delta}~log~sSFR_{\rm UV,cor}=-0.45$. 
Red lines indicate $\rm {\Delta}~log~sSFR_{\rm UV,cor}=-1.2$. 
Three boundary lines are combined to divide our galaxies into 
four sub-groups: SFGs above and below the SFMS ridge (SFMS-upper \& 
SFMS-lower), transition galaxies and quiescent galaxies. 
Four sub-groups are shown with dots in different colors. 
\label{sample}}
\end{figure*}

%------------------------------------------------------------------------------

Table 1 details our selection criteria and the resulting sample 
sizes after each cut for each field. A detailed discussion on sample completeness by these criteria 
is given in Appendix. 
After the cuts 1-5, we select 5545 galaxies in total from 
all five CANDELS fields. Furthermore, 4377 galaxies remain after the sixth cut, 
of which we utilized their spectroscopic redshifts for 1132 ($\sim25.86\%$) galaxies, 
grism redshifts for 2152 ($\sim49.17\%$) galaxies and photometric redshifts 
for 1093 ($\sim24.97\%$) galaxies.
Figure 1 shows the distributions of galaxies on the $UVJ$ and $UVI$ planes 
for total sample and a nearly face-on ($q>0.5$) subsample 
after the cuts 1-5, respectively. It can be seen that $UVI$ reproduces all the main features of $UVJ$, 
including the quenched region and the distinctive stripe patterns of sSFR, which 
is in agreement with the results initially presented by \citet[][]{Wang2017}. 
In Appendix, we show that these main features on $UVJ$ and $UVI$ diagrams are 
still strong for UV+IR rates, which strengths our analysis in this work.
However, given that SFRs involving IR data are subject to their own set of systematic biases (see Appendix), 
it is reasonable to adopt the UV-based rates.

In Figure 2, we show the sSFR-mass relation for our final sample of galaxies. 
Recent works \citep[e.g.,][]{Whitaker14,Lee2015} showed that 
the SFMS is not consistent with a single power-law at $0.5<z<2.5$. 
Instead, it is better fitted by a broken power-law, such that 
below a characteristic mass ($M_c$) of $M_{c} \sim 10^{10.2}M_{\odot}$, 
the SFMS has a redshift-independent slope of $\alpha \sim 1.0$. 
Above $M_c$, the SFMS has a shallow slope. 
At $0.5<z<1$, the characteristic mass is 
$\rm log(M_{c}/M_{\odot})=10.0\pm0.1$ \citep[][]{Whitaker14,Tomczak2016}.
Such the feature is obvious in our data.
To derive the SFMS relation of our galaxies, an initial fit to all $UVJ$-defined SFGs is made; objects more
than 2$\sigma$ away from the fit are then excluded for the next fit. This fitting process is repeated
until no new objects are excluded.
The best broken power-law fits to the SFMS of our galaxies are shown as black lines in Figure 2, 
and are described by the following equations: 
$\rm \log~sSFR_{\rm UV,cor}/yr^{-1}=0.02\pm0.02[\log~M_{\ast}/M_{\odot}-10]-8.91\pm0.12$ 
for $M_{\ast} \leq 10^{10}M_{\odot},$ and 
$\rm \log~sSFR_{\rm UV,cor}/yr^{-1}=-0.41\pm0.03[\log~M_{\ast}/M_{\odot}-10]-8.91\pm0.12$ 
for $M_{\ast} > 10^{10}M_{\odot}$ 

To quantify the relative star formation activity in galaxies in a given 
mass bin, we compute the vertical offsets in $\rm log~sSFR_{UV,cor}$ 
from the best-fit SFMS. The offset for a given galaxy 
is denoted by $\rm {\Delta}~log~sSFR_{UV,cor}$. 
Galaxies lying above (below) the best-fit SFMS are defined to 
have positive (negative) residuals. We then adopt $\rm {\Delta}~log~sSFR_{UV,cor}$ 
to divide our galaxies into the following four sub-groups: \\
a. $\rm {\Delta}~log~sSFR_{\rm UV,cor} \geq 0$ -- SFGs above the SFMS ridge, \\
b. $\rm -0.45~dex \leq {\Delta}~log~sSFR_{\rm UV,cor}<0$ -- SFGs below the SFMS ridge, \\
c. $\rm -1.2~dex \leq {\Delta}~log~sSFR_{\rm UV,cor}<-0.45~dex$ -- transition galaxies, \\
d. $\rm {\Delta}~log~sSFR_{UV,cor}<-1.2~dex$ -- quiescent galaxies. \\
The use of this relative quantity ($\rm {\Delta}~log~sSFR_{UV,cor}$) for 
classification means that our results are insensitive to the exact zero points and slopes of the fits. 
  
The SFMS has a dispersion of $\rm {\sigma}\sim0.3~dex$ in the logarithmic scale. 
The two thresholds of $\rm {\Delta}~log~sSFR_{UV,cor}={-0.45~dex}$ (blue lines) and 
$\rm {\Delta}~log~sSFR_{UV,cor}={-1.2~dex}$ (red lines) correspond to $1.5\sigma$ and $4\sigma$ below 
the ridge of the SFMS, respectively. 
We have checked that a change of $\rm \pm0.1 dex$ for our thresholds does not affect our conclusions. 

\section{Results and Analysis}

Following the work of \citet[][]{Wang2017}, in Figure 3 we show 
raw stacked $UVI$ color trajectories in $UVI$-space 
for four sub-groups of total sample galaxies in four stellar mass bins, 
$9.0<\log~M_{\ast}/M_{\odot}\leq9.5$, $9.5<\log~M_{\ast}/M_{\odot}\leq10.0$, 
$10.0<\log~M_{\ast}/M_{\odot}\leq10.5$ and $\log~M_{\ast}/M_{\odot}>10.5$, respectively. 
Color profiles in the same mass bins are stacked 
by first normalizing the radial positions of each galaxy by 
its $R_{SMA}$ and then computing the median colors 
at the selected normalized positions. 
In each panel, five radial locations are indicated by solid circles with increasing size, 
ranging from 0.2 $R_{SMA}$ at the innermost to 2.0 $R_{SMA}$ at the outermost. 
The arrows indicate the Calzetti reddening vector. 
The typical errors of the stacked colors are given for each panel, 
which are the average of the standard errors of median colors at all radii.
An example is given in Appendix to illustrate our estimate on the standard errors 
of median colors at a given radius.
%--------
In addition, we correct the PSF effects by adopting 
the method of \citet[][]{Szomoru2011}. 
The corrected light profiles are first generated 
by adding the residuals from fitting the raw light profiles 
with the PSF-convolved single S\'ersic model \citep[][]{Sersic1968} onto 
the S\'ersic profiles. 
The resultant color trajectories are overplotted as magenta solid curves 
with increasing radii indicated by an arrow in Figure 3. 
As can be seen, the PSF-correction stretches the length of color gradients 
on the $UVI$ diagram by making galaxy colors slightly redder in the centers 
and bluer in the outskirts. However, the overall effect on color gradients is not large, 
which is consistent with our evaluation in Appendix and that by \citet[][]{Wang2017}.
To minimize the interplay between stellar population and dust reddening changes, 
in Figure 4 we additionally show the $UVI$ color trajectories for a subsample 
of nearly face-on ($q>0.5$) galaxies. 
These two figures for the two samples are similar. Main features in common are listed as follows:

\begin{figure*}
\centering
\includegraphics[angle=0,width=1.0\textwidth]{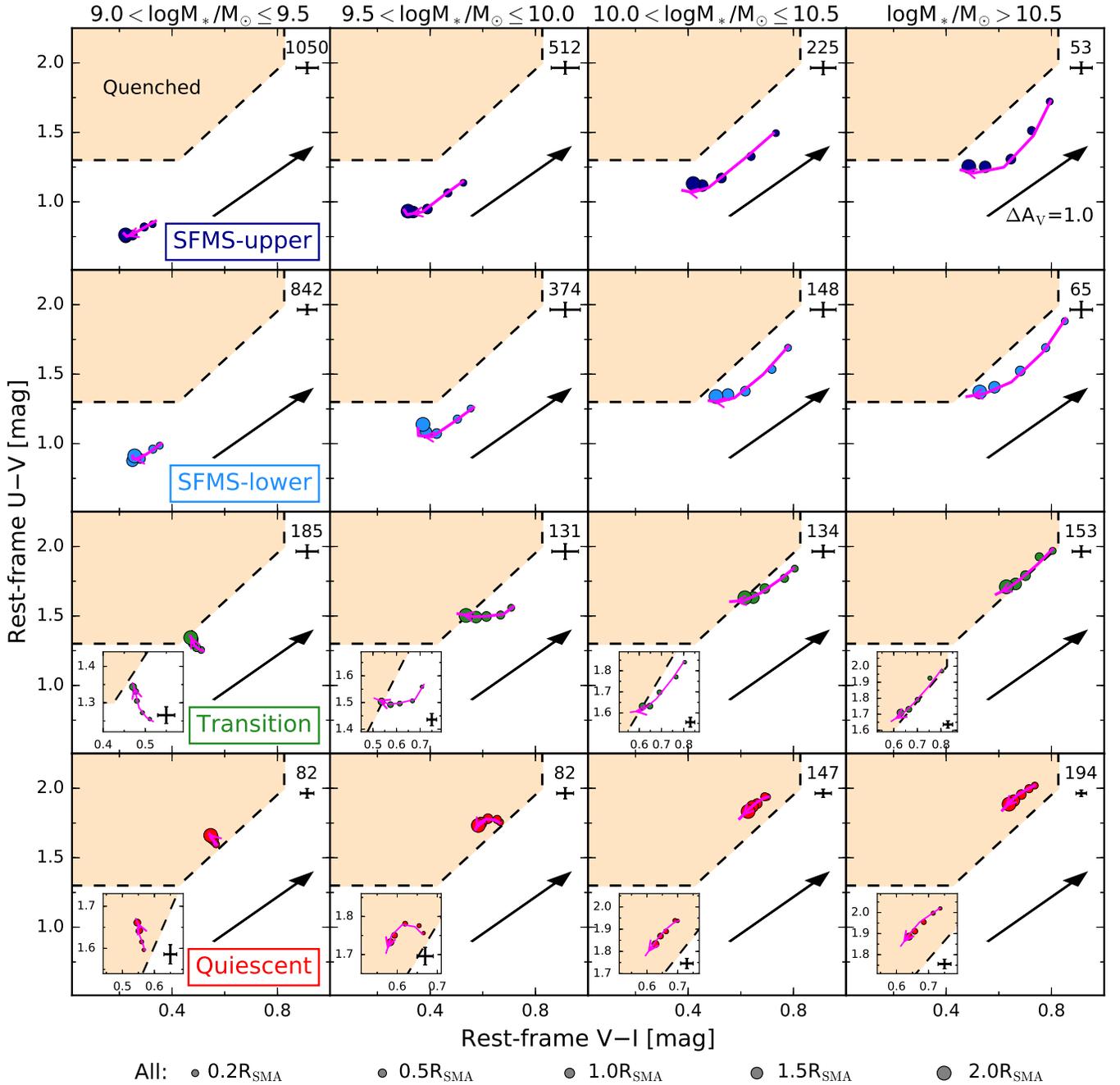}
\caption{
Rest-frame $UVI$ color gradients for all SFGs above and below the SFMS ridge (SFMS-upper \& SFMS-lower), 
transition galaxies and quiescent galaxies from top to bottom in
four mass bins respectively, as indicated at the top margin. For each panel, raw stacked data without correction for
PSF smearing are shown by solid circles with increasing size. Magenta solid curves with an arrow
indicate the radial trajectories after PSF correction, ranging from $\rm 0.2R_{SMA}$ to $\rm 2R_{SMA}$.
Dashed lines indicate the boundary to separate quiescent from star-forming galaxies. 
The arrows indicate the Calzetti reddening vector. The average standard deviations of the stacked colors 
are shown at upper right. The galaxy number in each panel is shown on the right-top corner as well.
An inserted zoom-in plot is given specifically 
for quiescent and transition galaxies to display the details on small scales. 
\label{uvi_gradient_all}}
\end{figure*}

\begin{figure*}
\centering
\includegraphics[angle=0,width=1.0\textwidth]{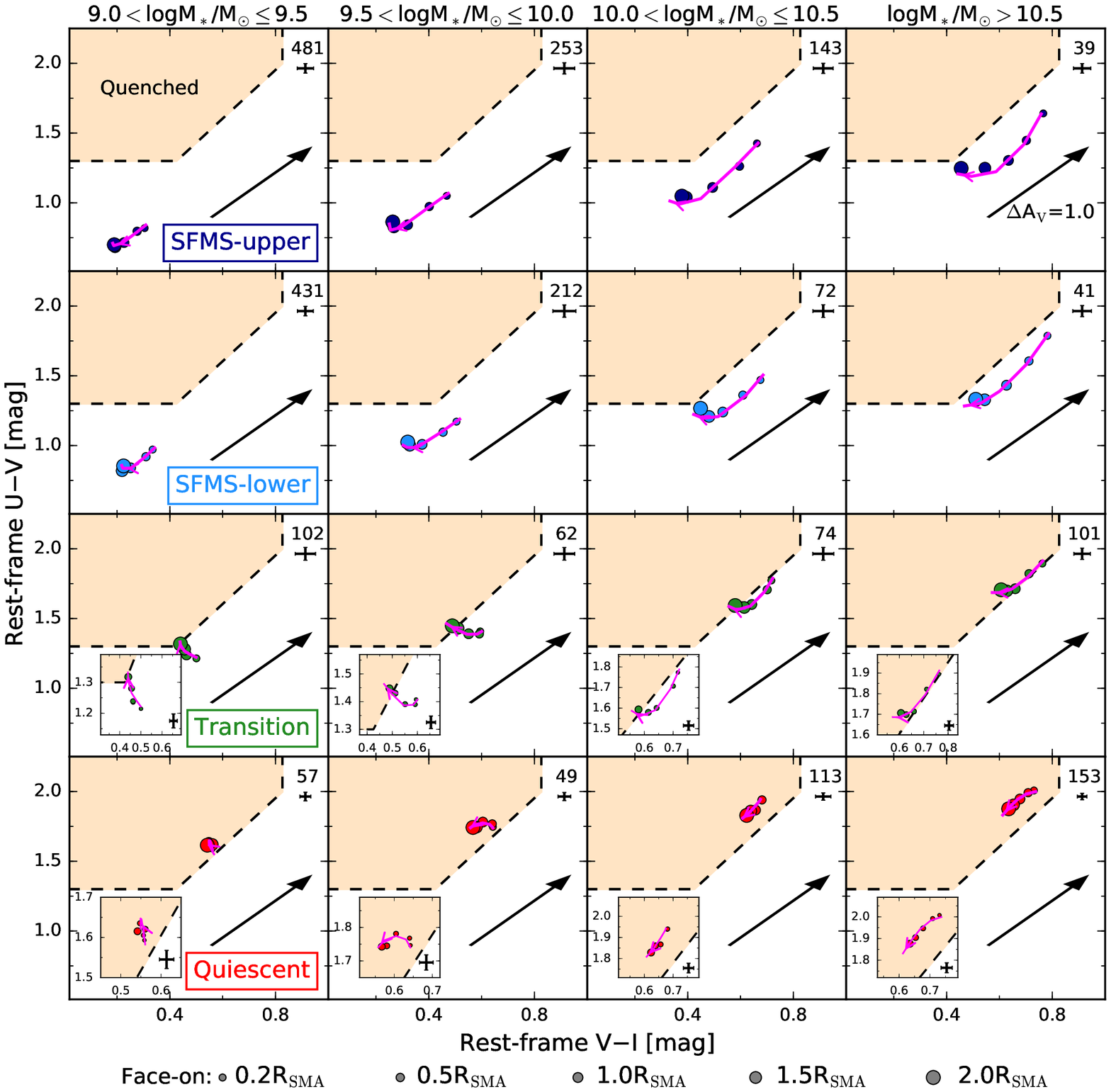}
\caption{
The same with Figure 3 but here only nearly face-on ($q>0.5$) galaxies are shown.
\label{uvi_gradient_face_on}}
\end{figure*}

\begin{enumerate}

\item For all mass ranges, all median color data points of SFGs both above and below the SFMS ridge lie well
within the star-forming region. For high-mass transition galaxies with $M_{\ast} > 10^{10}M_{\odot}$, 
all their median colors lie close to the boundary that
separates quenched from non-quenched galaxies. On the other hand, all median colors except for the ones in inner regions ($0.2R_{SMA}$) 
of low-mass ($M_{\ast} = 10^{9}-10^{10}M_{\odot}$) transition galaxies
lie still in the star-forming region of the $UVI$ diagram. The median colors of quiescent galaxies at all radii lie well within the quenched region for all mass ranges.  

\item The magnitudes of trajectories of SFGs are much larger than
those of transition and quiescent galaxies. This is likely due to
significant amounts of dust reddening in SFGs, because the magnitudes of their trajectories
are elongated along the direction of reddening vector.

\item The trajectories of SFGs both above and below the SFMS
are roughly parallel to the reddening vector, except that
the centers in most massive bins obviously deviate
toward the quenched region and the slight upturns appear
in the outermost parts near $\rm 2.0R_{SMA}$.
These features have been captured for the entire population of SFGs
and well studied by \citet[][]{Wang2017}.

\item The color trajectories of transition galaxies with $M_{\ast} > 10^{10}M_{\odot}$ and 
those with $M_{\ast} = 10^{9}-10^{10}M_{\odot}$ have different shapes, which indicates that 
the low-mass galaxies and high-mass galaxies have different sSFR gradients. 

\end{enumerate}

We now turn to the task of converting $UVI$ color trajectories into the sSFR profiles 
following the method of \citet[][]{Wang2017}. 
To create the maps of sSFR on the $UVI$ planes of Figure 1b and Figure 1d,
we first divide each distribution into multiple $0.05\times0.1$ mag rectangles.
In each rectangle we compute the median CANDELS
values of integrated sSFR and assign them to the center of the rectangle.
Then, given any position on each $UVI$ plane, the corresponding
sSFR values can be obtained by linearly interpolating among the nearby rectangle centers.
This method makes it possible to deduce sSFR values from $UVI$ with an $rms$ accuracy of
$\sim$0.15 dex \citep[see \S3 in][]{Wang2017}.
Adopting this calibration, the PSF-corrected $UVI$ color trajectories
are converted to radial sSFR profiles, as shown in Figure 5. 
It can be seen that, for lower mass bins ($M_{\ast} = 10^{9}-10^{10}M_{\odot}$), SFGs both above and
below the SFMS ridge generally have flat sSFR profiles,
whereas the transition galaxies in the same mass ranges generally have negative sSFR gradients
(sSFRs in the outskirts are more suppressed). In contrast, for the most massive 
bins ($M_{\ast} > 10^{10.5}M_{\odot}$),
SFGs above and below the SFMS ridge and transition galaxies generally
have varying degrees of positive sSFR gradients (more centrally-suppressed sSFRs relative to their outskirts). 
As expected, nearly flat or relatively weak sSFR gradients are observed in quiescent galaxies at all masses. 
The outside-in quenching for lower mass bins is indicated by the drop in sSFR of the transition galaxies, 
whereas the inside-out quenching for the most massive bins is best indicated by the rising sSFR of the lower SFMS galaxies. 
The results from inferred sSFR are in good agreement with those from the directly observed data of $UVI$ color gradients.

\begin{figure*}
\centering
\includegraphics[angle=0,width=0.95\textwidth]{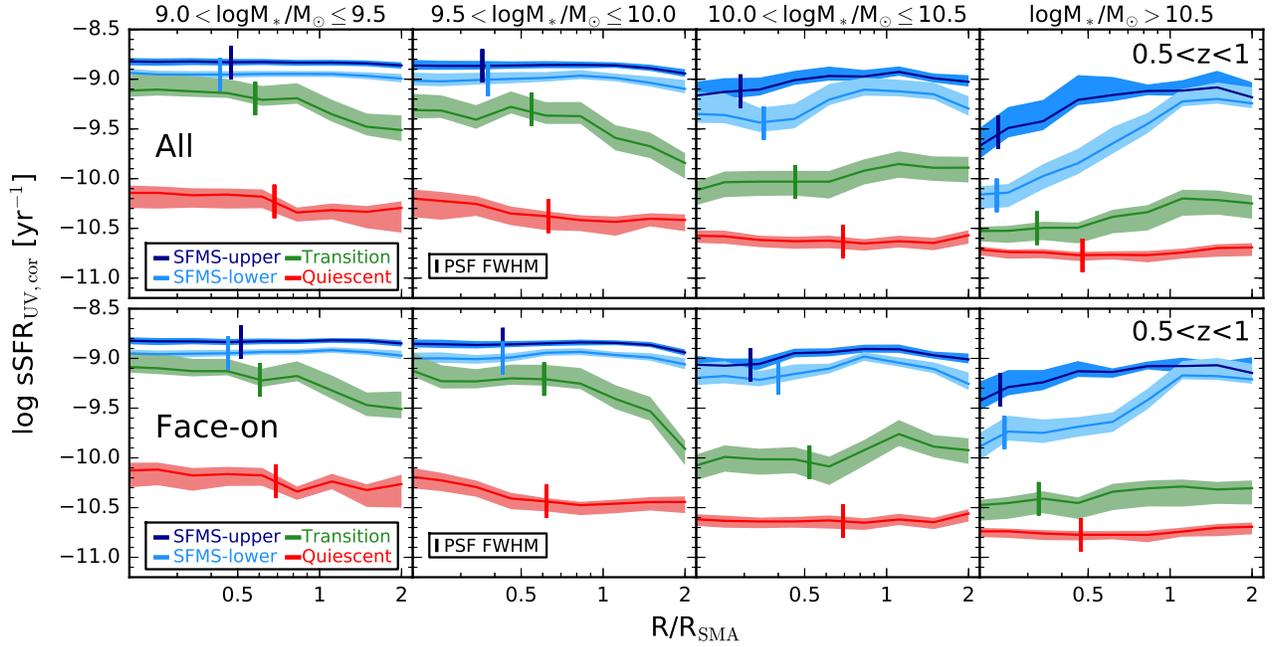}
\caption{
The stacked sSFR profiles of our total sample (top) and a nearly face-on subsample (bottom), 
which are inferred from the PSF-corrected $UVI$ color trajectories 
as in Figure 3 and 4, respectively. Radial distance is scaled 
by the median angular size $\rm R_{SMA}$ of each sub-group in each mass bin. 
Curves in different colors denote different sub-groups. Mass ranges are annotated 
at the top. Colored shadows denote the standard uncertainty of the stacked profiles, which 
include the uncertainty of the PSF correction and the standard deviation of the 
stacked color trajectories. 
Short vertical lines indicate the median PSF FWHM ($0.18^{\prime\prime}$) of each sub-group.
The outside-in quenching for lower mass bins is indicated by the drop in sSFR of the transition galaxies in green, 
whereas the inside-out quenching for the most massive bins is best indicated by the rising sSFR of 
the lower SFMS galaxies in blue.
\label{ssfr_profiles}}
\end{figure*}

\section{Discussion and Conclusion}

We select a sample of 4377 large ($R_{\rm SMA}>0.18^{\prime\prime}$) galaxies with $M_{\ast} > 10^{9}M_{\odot}$ 
between redshift 0.5 and 1.0 in all five CANDELS fields. 
These galaxies have well-measured, {\it HST} multi-band multi-aperture photometry data 
within $R < 2.0R_{SMA}$ in CANDELS. 
We investigate the stacked $UVI$ color gradients and inferred sSFR gradients 
in various galaxy populations
(quiescent galaxies, transition galaxies, and star-forming galaxies 
divided into above and below the ridge of the SFMS) in different mass bins.

We show that, for all mass ranges, star-forming galaxies, \emph{on average},
are not fully quenched at any radii, above the resolution limit of $0.18^{\prime\prime}$ ($\sim$1.3kpc), whereas
quiescent galaxies are fully quenched at all radii above this limit for all mass ranges. At all radii, the median UVI colors of high-mass transition galaxies with $\rm M_{\ast} > 10^{10}M_{\odot}$ lie close to the boundary that
separates quenched from non-quenched galaxies in the UVI diagram. In contrast,  all median colors except for the ones in their inner regions ($0.2R_{SMA}$) of low-mass ($M_{\ast} = 10^{9}-10^{10}M_{\odot}$) transition galaxies
lie still in the star-forming region of the $UVI$ diagram.

At low masses ($M_{\ast} = 10^{9}-10^{10}M_{\odot}$), SFGs both above and 
below the SFMS ridge generally have flat sSFR profiles, 
whereas the transition galaxies at the same masses have sSFRs that 
are more suppressed in the outskirts. In contrast, at high masses ($M_{\ast} > 10^{10.5}M_{\odot}$), 
SFGs above and below the SFMS ridge and transition galaxies 
have varying degrees of more centrally-suppressed sSFRs relative to their outskirts. 
These findings indicate that at $z\sim~0.5-1.0$ the main galaxy quenching mode depends on
its already formed stellar mass, exhibiting a transition from ``the outside-in''
at $M_{\ast} \leq 10^{10}M_{\odot}$ to ``the inside-out'' at $M_{\ast} > 10^{10.5}M_{\odot}$. 
The sSFR profiles in massive galaxies start to vary when they are on the SFMS,
whereas the sSFR profiles in the lower-mass galaxies start to vary when they move off the SFMS.
Similar trend is also observed in local galaxies \citep[][]{Perez2013,Pan15,Belfiore2017}. 
This pattern is broadly consistent with the prediction 
of \citet[][]{Tacchella06a,Tacchella06b} in cosmological simulations that 
a transition from outside-in to inside-out quenching occurs near a critical mass, 
$M_{\ast} \sim 10^{10}M_{\odot}$.
Our results support that the internal processes (i.e., central compaction, AGN feedback and supernova feedback) 
dominate the quenching of massive galaxies, whereas the external processes (i.e.,
environmental effects) dominate the quenching of low-mass galaxies.

We have checked that our main results are unchanged under the stacking
by physical radius rather than scaled radius.
We stress that the inferred sSFR gradients are in good agreement with
our raw data of $UVI$ color gradients. The former, however, depends
on the conventional SED modeling assumptions (i.e., $\tau$-models, solar metallicity
and a foreground-screen Calzetti reddening law).
We refer the reader to \citet[][]{Wang2017} for discussions about the effects of
these assumptions on sSFR gradients in SFGs (see \S8.2 and Appendix in their paper).
The Wang et al.'s critique shows that, as long as stellar populations are reasonably uniform throughout a galaxy,
these assumptions do not significantly affect the gradients in SFGs.
It is still unclear what biases these assumptions can bring to the inferred sSFR gradients
in transition and quiescent galaxies. 
The star formation histories of these populations are different from that of star-forming galaxies. 
Nevertheless, the resulting sSFR gradients in these populations based on the standard assumptions 
are consistent with the observed UVI color gradients, which are independent of any assumptions.

We caution that there are important ambiguities for highly-quenched populations, 
due to the degeneracies among the effects of dust, stellar age and metallicity on color gradients. 
Therefore, our conclusions for these objects are tentative.
When we convert their color trajectories to sSFR gradients, we find no significant trend
for the most massive bins. 
If these objects contain little or no dust \citep[see Figure 11 in][]{Fang2017} and 
their age (sSFR) gradients are flat, as shown in Figure 5, metallicity effects
would play an important role. This is quite similar to the origin of optical color gradients in nearby early-type galaxies \citep[e.g.,][]{Wu2005}.
In contrast, for lower mass bins we indeed see a trend that the centers are slightly
younger than the outer parts. This trend is indicated by a clear transverse motion of
the gradients across loci of constant sSFR.
Overall, the resulting quiescent trend agrees with a similar trend in transition galaxies 
at the same masses, suggesting that the two classes of galaxies are evolutionarily linked, as
we would expect, since the mass difference between transition and quiescent galaxies should be small.
We also caution that perhaps gradients in lower-mass quiescent galaxies
cannot be resolved simply because they are too small.
Future works should investigate the consequences of more realistic stellar population models, 
metallicity, and dust extinction law.

\acknowledgments

We acknowledge the anonymous referee for constructive comments and suggestions 
that significantly improved this paper.
We thank Steven Willner for helpful comments on the manuscript, and Shude Mao for 
useful discussion. We acknowledge supports from the NSF grants of China (11573017,11733006) 
and the CANDELS program HST-GO-12060 by NASA through a grant from the STScI. 
Space Telescope Science Institute, which is operated by the Association of Universities 
for Research in Astronomy, Incorporated, under NASA contract NAS5-26555. 
This work is partly based on observations taken by the 3D-HST Treasury Program (GO 12177 and 12328) 
with the NASA/ESA HST, which is operated by the Association of Universities 
for Research in Astronomy, Inc., under NASA contract NAS5-26555.
S.M.F., D.C.K., Y.G., and H.M.Y. acknowledge partial support from US NSF grant AST-16-15730. 
P.G.P.-G. acknowledges support from Spanish Government MINECO AYA2015-70815-ERC 
and AYA2015-63650-P Grants.

%%-------------

%\bibliographystyle{aasjournal}
%\bibliography{journals}

\clearpage

\appendix

\section{$\lowercase{s}SFR_{\rm UV+IR}$ patterns on $UVJ$ and $UVI$ Planes}

We follow the \citet[][]{Rujopakarn2013ApJ} method to derive $L_{\rm IR}$ of our sample galaxies. 
Only F160W objects that are identified as the nearest neighbor to a MIPS source are retained. 
As a result, of 5545 galaxies satisfying our selection criteria 1-5, 3790 ($\sim 70\%$) have $S/N>1$ MIPS $24{\mu}m$ 
detections, 957 are undetected in MIPS $24{\mu}m$ photometry ($S/N<1$), 782 have negative MIPS $24{\mu}m$ fluxes, 
and 16 are unmatched.
We then use the formula presented in \citet[][]{Wuyts2011ApJ} to calculate the UV+IR SFRs for MIPS-detected subsample:
\begin{equation}
{\rm SFR_{UV+IR}} [M_{\odot}yr^{-1}]=1.09\times10^{-10}(L{\rm_{IR}+3.33}L{\rm_{UV}}) [L_{\odot}]
\label{1}
\end{equation}
where $L_{\rm IR}$ is the integrated 8$-$1000 $\mu$m luminosity inferred from MIPS $24{\mu}m$, and
L${\rm_{UV}}\equiv\nu L_{\nu}(2800\AA)$
is the rest-frame near-UV luminosity, measured at 2800\AA.
In Figure 6, we re-plot the sSFR-mass relation of all galaxies (left) and make a direct comparison
between $sSFR_{\rm UV,cor}$ and $sSFR_{\rm UV+IR}$ for $24{\mu}m$-detected objects only (right).
As can be seen, in our redshift range, the UV+IR rates are biased at lower masses and for lower-sSFR objects. 
The $UV-$based method tends to slightly overestimate the SFRs for high-sSFR galaxies, which is in agreement with 
the assessment of \citet[][]{Fang2017}.
For lower-sSFR objects (i.e., transition and quiescent galaxies), $sSFR_{\rm UV+IR}$ are systematically
higher than $sSFR_{\rm UV,cor}$. This inconsistency likely originates from the uncertainty of MIPS photometry
(i.e., the deepest IR data available in GOODS-S and GOODS-N gives better consistency than the other
three fields) or it is because the $24{\mu}m$ flux in these objects comes, at least in part, from sources other than
dust heated by conventional star formation \citep[i.e., from old stars, see][and reference therein for discussions]{Fang2017}.
For these reasons, it is reasonable to adopt the UV-based rates
especially when sSFR is low, as we do in this paper.
In Figure 7, we re-plot the $UVJ$ and $UVI$ planes for the MIPS $24{\mu}m$ detected subsample only, this time 
galaxies are color-coded by their $\rm log~sSFR_{UV+IR}$. Each individual is color-coded in the top two panels, whereas the 
median value in each bin is color-coded in the bottom two panels, which contains at least 5 objects.
It can be seen that statistically the main features, including the quenched region
and the distinctive stripe patterns of sSFR, are still strong when using independent UV+IR
SFRs. This is in agreement with the result of \citet[][Figure 25]{Straatman16}, 
which strengthens our analysis in this work.

\section{Sample Completeness}

We discuss the resulting sample completeness by our selection criteria. 
First, the criterion $H_{mag}<24.5$ (criterion 1) can select relatively complete mass-limited samples 
of both blue and red galaxies above $\rm M_{\ast} \sim 10^{9}M_{\odot}$ in the redshift range 
$z=0.5-1$ (criterion 3) \citep[see Figure 2 in][]{vdWel14}. 
Second, {\tt SExtractor} parameter $\rm CLASS\_STAR$ cut (criterion 2) is as powerful as 
colors to separate galaxies from stars \citep[see Figure 18 in][]{Guo+13}.
Of particular notes are the criteria 4-6. The {\tt GALFIT} $\rm flag$ cut (criterion 4) 
excludes $\approx4\%$ of sample galaxies after the cuts 1-3.  
Visual inspection shows that these galaxies discarded by this cut are either mergers or strongly contaminated by 
neighbor objects. The multi-aperture photometry on these objects by using the {\tt IRAF} routine {\tt ellipse} 
usually fails. 
Furthermore, to minimize the PSF effects on color measurement (see below), we applied the 
angular size cut, $R_{\rm SMA}>{\rm 3~drizzled~pixels}$ (criterion 5). 
In Figure 8, we show the $UVJ$ diagram, sSFR-mass relation and size-mass relation for
both large and small galaxies after the cuts 1-4. As seen from the plots, 
the angular size cut preferentially removes more 
low-mass galaxies below $\rm M_{\ast} \sim 10^{10.5}M_{\odot}$. This lack of sample completeness
in this work cannot be avoided given the limited available resolution of $\it HST$ images \citep[][]{Wang2017}, 
otherwise the resulting color gradients of very small galaxies are likely artificial.
In Figure 9, we show that, in the sSFR-mass space,
the distribution of galaxies satisfying the criteria 1-6 is quite similar to that satisfying the criteria 1-5,
which indicates that the criterion 6 does not create significant bias for
this analysis.
Finally, to enable the reader to know which bin is impacted the most, in Table 2 we specifically provide 
the resulting sample sizes by each cut after the third criterion for each sub-group in different 
mass bins. 

\section{PSF Effects on the measurement of Color Gradients}

In the documentation of the CANDELS {\it HST} multi-band and multi-aperture photometry catalogs 
still under construction by Liu et al. (in preparation), we will make a detailed assessment on the PSF effects. 
So far the effects of PSF mis-matching and PSF smearing on derived color gradients have been evaluated as below.

\subsection{Effect of the PSF mis-matching}

Stars should have no observed color gradients, so the color difference
between any two observed bands should be zero at all radii.
To check this, we have carefully selected some unsaturated stars brighter than 24 magnitude 
with $\rm CLASS\_STAR > 0.95$ in the GOODS-S field. 
Figure 10 shows their integrated color differences between $H(F160W)$ and bluer bands 
($J(F125W)$, $I(F814W)$ and $V(F606W)$). 
The result indicates that $V(F606W)$ and $I(F814W)$ are under-smoothed so the centers of the stars
are bluer than the total colors (i.e., artificially bluer
than what they should be). $J(F125W)$ is over-smoothed so the centers are redder there.
These trends are similar to those in the GOODS-S photometry paper
\citep[][]{Guo+13}.
This is probably the best we can do with IRAF/PSFMATCH program.
At $R>3$ pixels ($0.18^{\prime\prime}$), the median deviations (red circles) 
in $I(F814W)$ and $J(F125W)$ are almost zero. The deviation in $V(F606W)$ is larger,
but still less than $\sim$0.025 mag.

\subsection{Effect of the PSF smearing}

Besides the mis-matching issue among different bands addressed above, the effect of PSF smearing can also
make the observed color gradients different from the intrinsic ones, especially for galaxies with small angular sizes 
and the galaxy central regions. 
We systematically evaluate this effect by modeling mock galaxies with various shapes and color gradients,
and then convolve them with the F160W PSF to obtain smeared images, as we did for realistic galaxies. 
The color gradients of these output images are then compared with the intrinsic values.
We assume that mock galaxies in one red band have ideal single-component S\'ersic profiles,
with $R_{\rm SMA}$=0.2, 0.35, 0.5, 0.65 arcsec (note that the drizzled imaging
pixel scale is 0.06 arcsec/pixel), S\'ersic index $n$=0.5, 1.0, 1.5, 2.0
and ellipticity $\epsilon$=0.0, 0.3, 0.5, 0.7
well represent the majority of realistic galaxies.
Then one of logarithmic color gradients G = d color/d log r = -0.1, -0.3, -0.5
is superimposed onto the S\'ersic profile, except that
a linear component is adopted for central pixel instead to avoid logarithmic divergence,
as the image of the other bluer band. 
The observed normalized color gradients are then compared to the unsmeared unit color gradient.

In Figure 11, we show the results for these mock galaxies with
typical shapes and gradient (median values). Zero points are all
fixed at $R_{\rm SMA}$. It can be seen that the deviation is fast enlarged after the radial distance
decreases to 3 pixel sizes ($0.18{\arcsec}$), 
suggesting that PSF smearing effect roughly diminishes at 0.18 arcsec away from galaxy centers. 
Therefore, it is reasonable to exclude small galaxies with effective radius less than $0.18{\arcsec}$ and
take the color of $R > 0.18{\arcsec}$ as a safe indicator of the intrinsic color. 
In conclusion, the overall PSF effects are not large 
for real data at $R > 0.18{\arcsec}$. 
The PSF-correction can result in an error of $\sim$0.05 in our color gradients at most
\citep[also see Figure 5 in][]{Wang2017}.

\section{Error estimate on the stacked color trajectories}

In Figure 3 and Figure 4, we provide the typical errors of the stacked colors in each panel, 
which are the average of the standard errors of median colors at all radii. 
Figure 12 presents an example to illustrate our estimate on the standard errors of the median 
colors at a given radius. For the $UVI$ colors at $R_{SMA}$ of face-on SFMS-upper galaxies in the mass bin
$M_{\ast} = 10^{10}-10^{10.5}M_{\odot}$, the photometric errors (i.e., readout noise, sky subtraction, 
and PSF matching, etc.) of individual data points are typically $0.1-0.15$ magnitudes (left). 
The standard error of the median $U-V$ or $V-I$ colors at this radius is 
computed using a Monte Carlo simulation to re-generate 1000 realizations of 
median colors after resampling at each iteration of the errors 
of each data point from Gaussian distribution, with zero means and standard deviations 
given by the observed photometric errors. 
The resulting 1000 median colors are assumed to follow a Gaussian distribution. 
The standard error of the median colors at this radius is roughly the same as the
standard deviation of the assumed Gaussian distribution (right).

%%-------------
\begin{figure*}
\centering
\includegraphics[angle=0,width=0.65\textwidth]{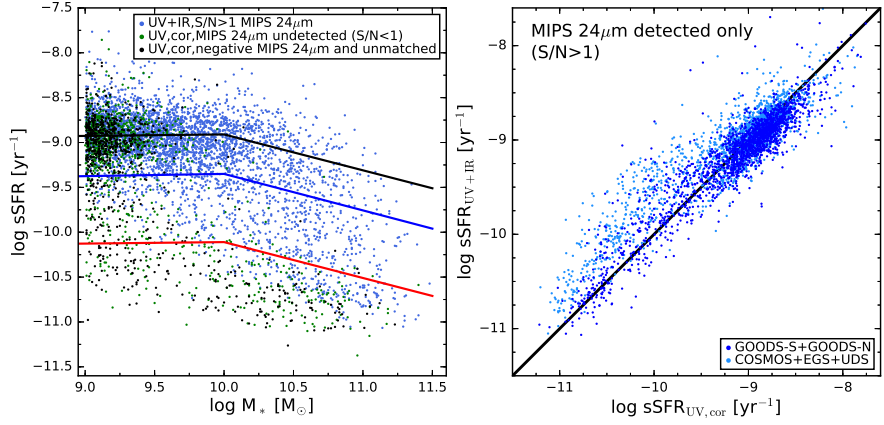}
\caption{
The sSFR-mass relation for all galaxies after the cuts 1-5 (left) and 
$sSFR_{\rm UV+IR}$ versus $sSFR_{\rm UV,cor}$ for MIPS-detected subsample only (right). 
In the left panel, blue dots denote galaxies with $S/N>1$ MIPS $24{\mu}m$ detections. 
Green dots denote galaxies undetected in MIPS $24{\mu}m$ photometry ($S/N<1$). 
Black dots denote galaxies that either have negative MIPS $24{\mu}m$ fluxes or are unmatched sources. 
Solid lines have the same meanings as those in Figure 2. In the right panel, 
galaxies with the deepest IR data in GOODS-S and GOODS-N and galaxies in the 
other three fields (COSMOS, EGS and UDS) are shown in different colors.
\label{calibration}}
\end{figure*}

\begin{figure*}
\centering
\includegraphics[angle=0,width=0.65\textwidth]{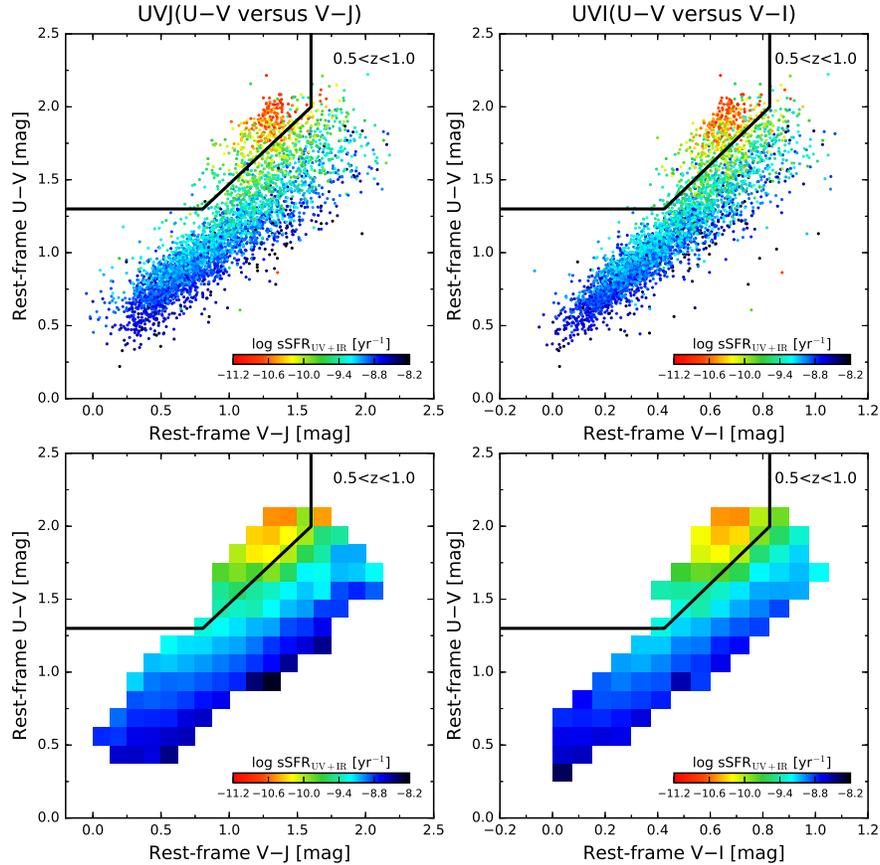}
\caption{
Rest-frame global $UVJ$ diagrams (left) and $UVI$ diagrams (right) for
the MIPS $24{\mu}m$ detected subsample after the cuts 1-5, which are
color-coded by $\rm log~sSFR_{UV+IR}$.
Each individual is color-coded in the top two panels. In contrast,
the median value in each bin is color-coded in the bottom two panels,
which contains at least 5 objects. Solid lines have the same meanings as those in Figure 1.
\label{calibration}}
\end{figure*}

\begin{figure*}
\centering
\includegraphics[angle=0,width=0.95\textwidth]{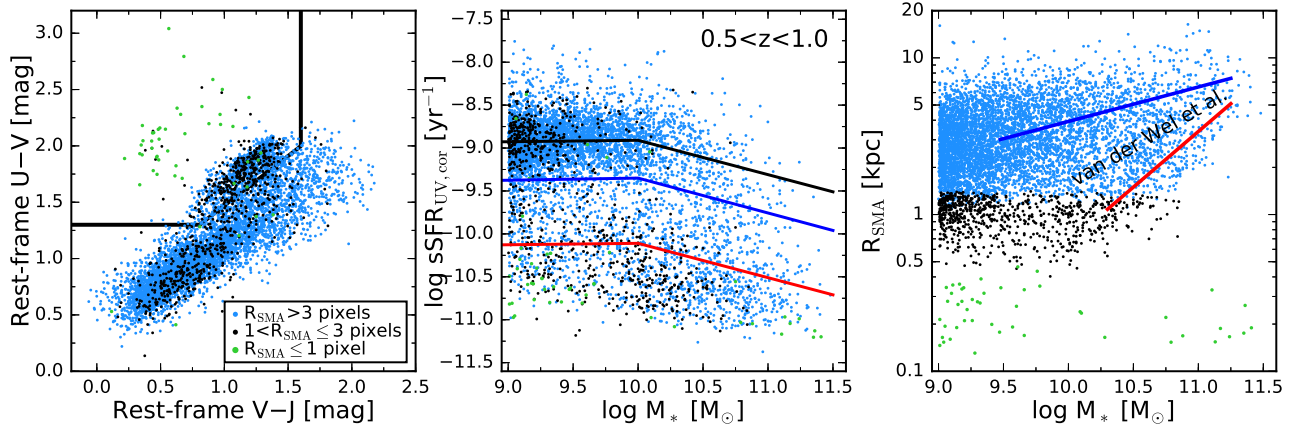}
\caption{
Rest-frame $UVJ$ diagram (left), sSFR-mass relation (middle) and size-mass relation (right)
for all galaxies after the cuts 1-4. Galaxies are color-coded by their effective radii.
Solid lines in the left panel have the same meanings as those in Figure 1. 
Solid lines in the middle panel have the same meanings as those in Figure 2.
Solid lines in the right panel show the best-fit size-mass relations for 
star-forming galaxies with $M_{\ast} > 3\times10^{9}M_{\odot}$ (blue) 
and quiescent galaxies with $M_{\ast} > 2\times10^{10}M_{\odot}$ (red) determined by \citet[][]{vdWel14}, respectively.
Note that no additional corrections on effective radii are applied here \citep[see][for details]{vdWel14}.  
\label{calibration}}
\end{figure*}

\begin{figure*}
\centering
\includegraphics[angle=0,width=0.95\textwidth]{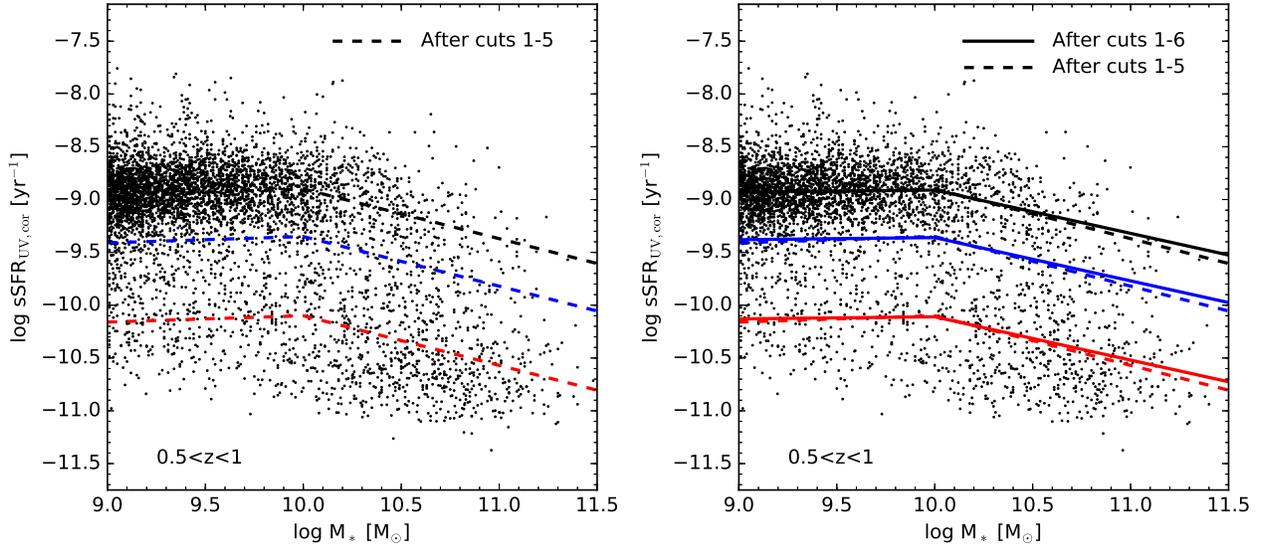}
\caption{
The sSFR-mass relations for our sample of galaxies after the cuts 1-5 (left) and that 
after the cuts 1-6 (right). Black lines are the best broken power-law fits to the SFMS ($\rm {\Delta}~log~sSFR_{\rm UV,cor}=0$). 
Blue lines indicate $\rm {\Delta}~log~sSFR_{\rm UV,cor}=-0.45$. 
Red lines indicate $\rm {\Delta}~log~sSFR_{\rm UV,cor}=-1.2$. 
As can be seen, the distribution of galaxies satisfying the cuts 1-6 is quite similar 
to that satisfying the cuts 1-5. 
The best-fit SFMS of galaxies after the cuts 1-5 (dashed lines) can be described by the following equations: 
$\rm \log~sSFR_{\rm UV,cor}/yr^{-1}=0.06\pm0.03[\log~M_{\ast}/M_{\odot}-10]-8.90\pm0.13$
for $M_{\ast} \leq 10^{10}M_{\odot},$ and
$\rm \log~sSFR_{\rm UV,cor}/yr^{-1}=-0.47\pm0.04[\log~M_{\ast}/M_{\odot}-10]-8.90\pm0.13$
for $M_{\ast} > 10^{10}M_{\odot}$, which are only slightly different with 
the best-fit SFMS of our final sample (solid lines in the right panel, see \S3).
\label{calibration}}
\end{figure*}

\begin{figure*}
\centering
\includegraphics[angle=0,width=0.85\textwidth]{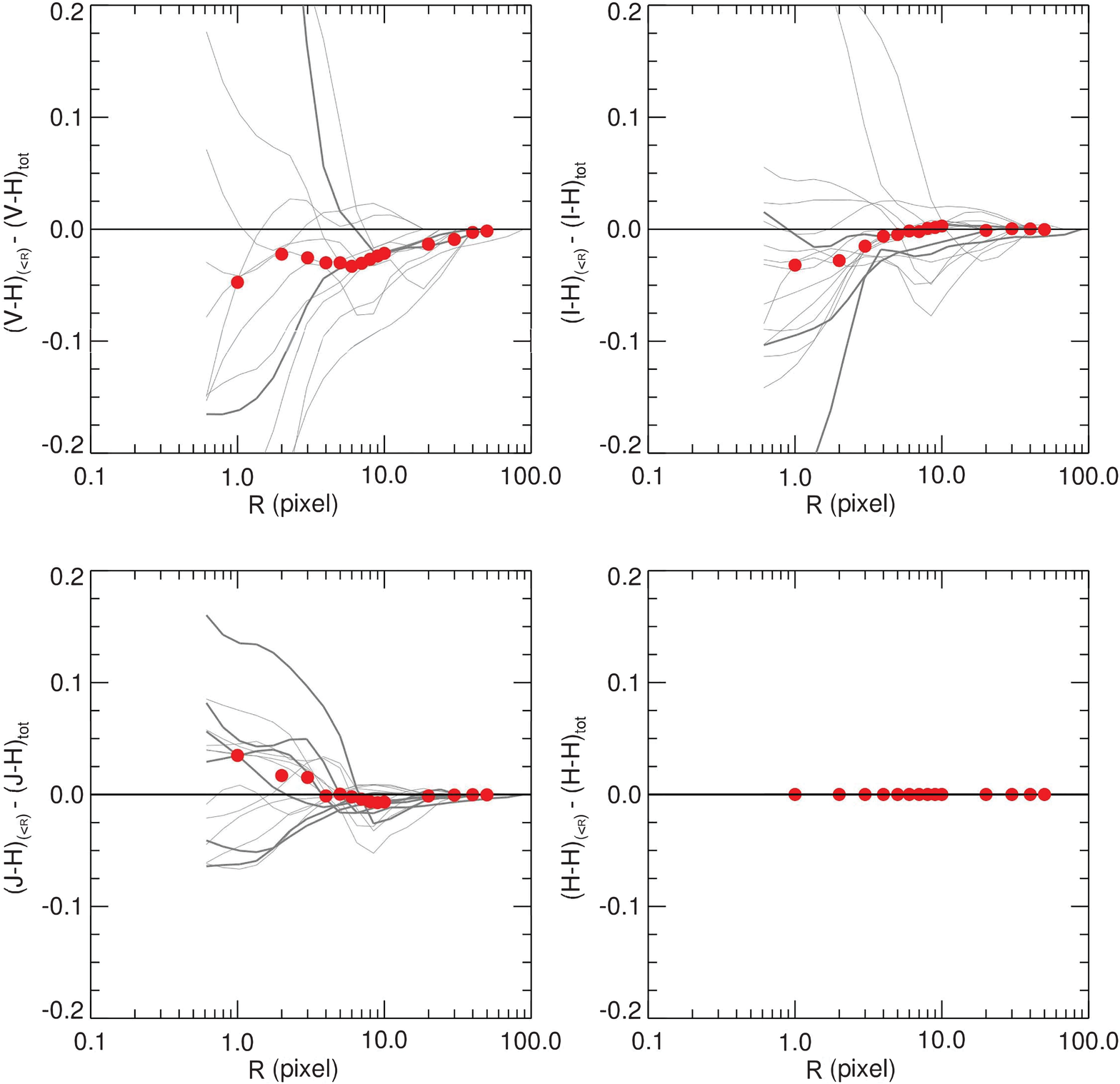}
\caption{
Integrated color differences between $H(F160W)$ and bluer bands ($J(F125W)$, $I(F814W)$ and $V(F606W)$)
for unsaturated stars brighter than 24 magnitude in $H(F160W)$ band in the GOODS-S field.
Gray lines show individual stars. Red circles show the medians.
Stars should have no observed color gradients, so the difference between any two bands
should be zero at all radius. This check shows that $V$ and $I$ are under-smoothed
so the centers of the stars are bluer than the total colors (i.e., artificially bluer
than what they should be). $J$ is over-smoothed so the centers are redder there. 
At $R>3$ pixels, the deviations of the red circles in $I$ and $J$ are almost zero.
The deviation in $V$ is larger, but still less than $\sim$0.025 mag.
\label{calibration}}
\end{figure*}

\begin{figure*}
\centering
\includegraphics[angle=0,width=0.62\textwidth]{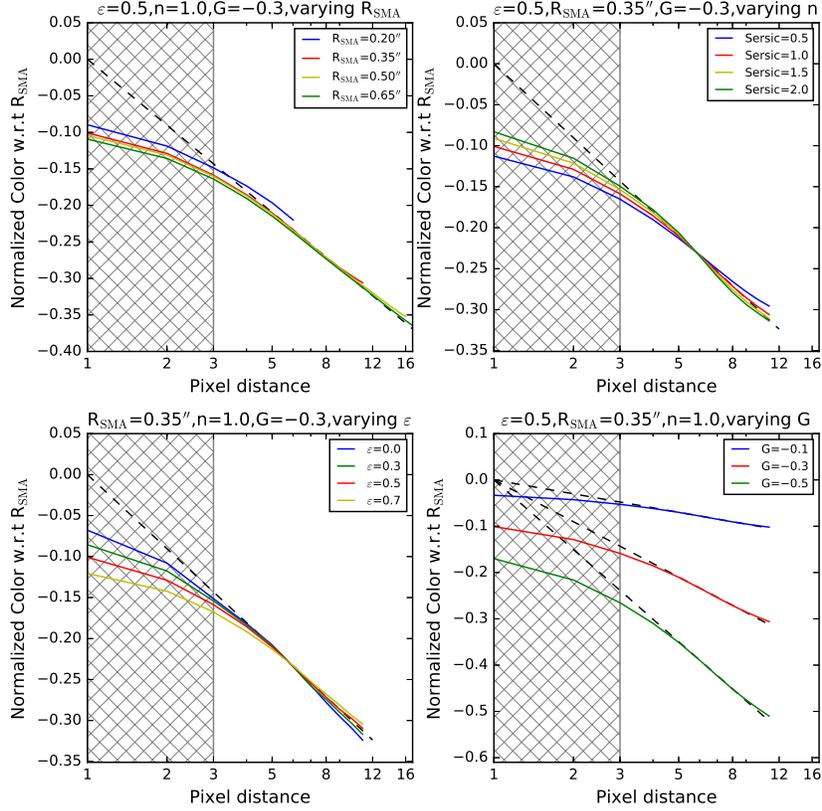}
\caption{
Normalized smeared color gradients in comparison with the intrinsic unit logarithmic
gradients ($G$) for mock galaxies with different $R_{\rm SMA}$, ellipticity ($\epsilon$) and S\'ersic index. Zero points are all fixed
at $R_{\rm SMA}$.
\label{calibration}}
\end{figure*}

\begin{figure*}
\centering
\includegraphics[angle=0,width=0.62\textwidth]{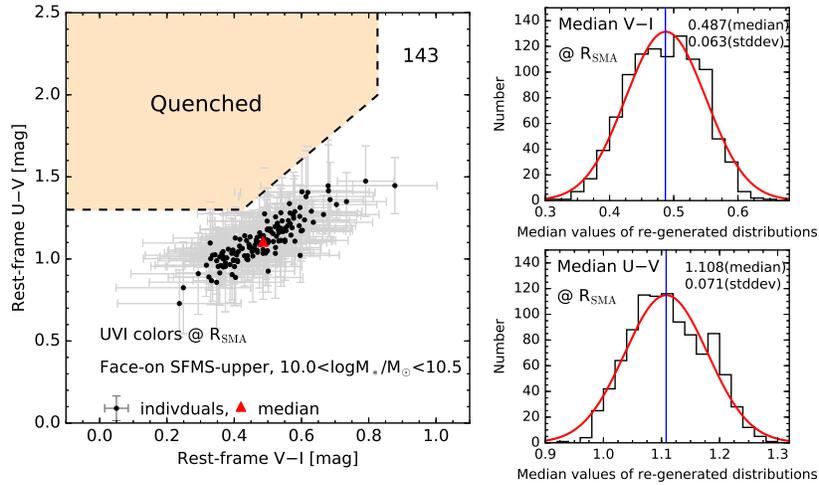}
\caption{
An example to illustrate our estimate on the standard errors of median colors at a given radius
in a given panel. 
Left panel shows the UVI color distribution at $R_{SMA}$ for face-on SFMS-upper galaxies in the mass bin
$M_{\ast} = 10^{10}-10^{10.5}M_{\odot}$. The observed photometric errors (i.e., readout noise, sky subtraction,
and PSF matching, etc.) of individual points are
given, which are typically $0.1-0.15$ magnitudes. Median values are shown with a red triangle.
The standard error of the median $U-V$ or $V-I$ colors at this radius is
computed using a Monte Carlo simulation to re-generate 1000 realizations of
median colors after resampling at each iteration of the errors
of each data point from Gaussian distribution, with zero means and standard deviations 
given by the observed photometric errors. 
Right panels show the distributions of these newly generated median U-V and V-I colors. 
The standard deviations are computed by assuming their distributions follow a Gaussian form.
The galaxy number in this panel is shown on the right-top corner.
\label{calibration}}
\end{figure*}

\clearpage

%-------------
\begin{table*}[ht]
\footnotesize
\begin{center}
\caption{The resulting sample sizes by each cut after the third criterion for each sub-group in different mass bins.}
\begin{tabular}{|c|c|c|c|c|}
\hline\hline
{\footnotesize Criteria} & {\footnotesize $9.0<\log~M_{\ast}/M_{\odot}\leq9.5$} &
{\footnotesize $9.5<\log~M_{\ast}/M_{\odot}\leq10.0$} & 
{\footnotesize $10.0<\log~M_{\ast}/M_{\odot}\leq10.5$} & 
{\footnotesize $\log~M_{\ast}/M_{\odot}>10.5$} \\
\hline
\multicolumn{5}{|c|}{\textbf{\footnotesize SFGs above the SFMS ridge (SFMS-upper)}} \\
\cline{1-5}
 1-3         &  1524(100\%)~~~    & 662(100\%)~~~      &  282(100\%)~~~      &   64(100\%)~~   \\
 1-4         &  1497(98.23\%)  & 651(98.34\%)    &  274(97.16\%)    &   63(98.44\%)  \\
 1-5         &  1319(86.55\%)  & 629(95.02\%)    &  269(95.39\%)    &   62(96.88\%)   \\
 1-6         &  1050(68.90\%)  & 512(77.34\%)    &  225(79.79\%)    &   53(82.81\%)   \\
\hline
\multicolumn{5}{|c|}{\textbf{\footnotesize SFGs below the SFMS ridge (SFMS-below)}} \\
\cline{1-5}
 1-3         &  1174(100\%)~~~    & 486(100\%)~~~      &  210(100\%)~~~      &   85(100\%)~~   \\
 1-4         &  1156(98.47\%)  & 478(98.35\%)    &  199(94.76\%)    &   82(96.47\%)  \\
 1-5         &  1036(88.25\%)  & 459(94.44\%)    &  194(92.38\%)    &   82(96.47\%)   \\
 1-6         &  842(71.72\%)   & 374(76.95\%)    &  148(70.48\%)    &   65(76.47\%)   \\
\hline
\multicolumn{5}{|c|}{\textbf{\footnotesize Transition galaxies}} \\
\cline{1-5}
 1-3         &  344(100\%)~~~     & 220(100\%)~~~      &  212(100\%)~~~      &   196(100\%)~~   \\
 1-4         &  326(94.77\%)   & 217(98.64\%)    &  199(93.87\%)    &   188(95.92\%)  \\
 1-5         &  258(75.00\%)   & 168(76.36\%)    &  174(82.08\%)    &   185(94.39\%)   \\
 1-6         &  185(53.78\%)   & 131(59.55\%)    &  134(63.21\%)    &   153(78.06\%)   \\
\hline
\multicolumn{5}{|c|}{\textbf{\footnotesize Quiescent galaxies}} \\
\cline{1-5}
 1-3         &  240(100\%)~~~     & 230(100\%)~~~      &  345(100\%)~~~      &   322(100\%)~~   \\
 1-4         &  204(85.00\%)   & 201(87.39\%)    &  325(94.20\%)    &   296(91.93\%)  \\
 1-5         &  117(48.75\%)   & 117(50.87\%)    &  212(61.45\%)    &   264(81.99\%)   \\
 1-6         &  82(34.17\%)    & 82(35.65\%)     &  147(42.61\%)    &   194(60.25\%)   \\
\hline
\end{tabular}
\end{center}
\end{table*}

%%-------------

\end{document}